\documentclass[10pt, conference, letterpaper]{IEEEtran}
\IEEEoverridecommandlockouts
\usepackage{graphicx}
\usepackage{subfigure}
\usepackage{cite}
\usepackage{booktabs}
\usepackage{amsmath,amssymb,amsfonts}
\usepackage{algorithmic}
\usepackage{graphicx}
\usepackage{textcomp}
\usepackage{makecell}
\usepackage{amsmath}
\usepackage{algorithm}
\usepackage{algorithmic}
\usepackage{multirow}
\usepackage{xcolor}
\usepackage{url}                      
\def\BibTeX{{\rm B\kern-.05em{\sc i\kern-.025em b}\kern-.08em
    T\kern-.1667em\lower.7ex\hbox{E}\kern-.125emX}}
\begin{document}

\title{BLADE: Behavior-Level Anomaly Detection \\ Using Network Traffic in Web Services}

\author{
    Zhibo Dong, 
    Yong Huang\textsuperscript{*},
    Shubao Sun,
    Wentao Cui,
    Zhihua Wang \\
    \textsuperscript{}School of Cyber Science and Engineering, Zhengzhou University, Zhengzhou 450001, China\\
    Email: \{dongzhibo, wentao\}@gs.zzu.edu.cn,  \{yonghuang, zhwang\}@zzu.edu.cn, sunzzu@stu.zzu.edu.cn\\
    \thanks{\textsuperscript{*}The corresponding author is Yong Huang (yonghuang@zzu.edu.cn).}
}

\maketitle

\begin{abstract}
With their widespread popularity, web services have become the main targets of various cyberattacks.
Existing traffic anomaly detection approaches focus on flow-level attacks, yet fail to recognize behavior-level attacks, which appear benign in individual flows but reveal malicious purpose using multiple network flows.
To transcend this limitation, we propose a novel unsupervised traffic anomaly detection system, BLADE, capable of detecting not only flow-level but also behavior-level attacks in web services. 
Our key observation is that application-layer operations of web services exhibit distinctive communication patterns at the network layer from a multi-flow perspective.
BLADE first exploits a flow autoencoder to learn a latent feature representation and calculates its reconstruction losses per flow.
Then, the latent representation is assigned a pseudo operation label using an unsupervised clustering method.
Next, an anomaly score is computed based on the reconstruction losses.
Finally, the triplets of timestamps, pseudo labels, and anomaly scores from multiple flows are aggregated and fed into a one-class classifier to characterize the behavior patterns of legitimate web operations, enabling the detection of flow-level and behavior-level anomalies.
BLADE is extensively evaluated on both the custom dataset and the CIC-IDS2017 dataset.
The experimental results demonstrate BLADE's superior performance, achieving high F1 scores of 0.9732 and 0.9801, respectively, on the two datasets, and outperforming traditional single-flow anomaly detection baselines.
\end{abstract}


\section{Introduction}

Web services play a pivotal role in today's digital landscape and enable seamless communication and data exchange between different applications and systems.
With their increasing prevalence, web services have become prime targets for a wide range of cyberattacks. 
It is reported that in 2024, attacks on web services have surpassed 311 billion and resulted in approximately 87 billion dollars in global losses~\cite{akamaistate2025}. 
Malicious traffic analysis is a critical component in safeguarding web services, enabling timely threat detection and mitigation~\cite{nascita2024survey, dong2025deep}.
Since malicious traffic represents a negligible share of the total network traffic and is evolving at a rapid pace, a popular strategy is to design malicious traffic analysis as an anomaly detection task~\cite{Mateenad, han2023anomaly, fu2024flow}.

Despite growing attempts and extensive endeavors, existing traffic anomaly detection approaches focus on verifying a single network flow and limit themselves to flow-level attacks such as injection, scanning, and brute-force attacks~\cite{fsnet, lstmae, min2021memae, qing2023low, liu2025decentralized}.
These attacks are often triggered by malicious payloads or abnormal communication patterns observable within a single network flow.
However, these approaches struggle to detect behavior-level attacks, which rarely reveal malicious intent within a single flow and instead exploit vulnerabilities of application-layer rules and configurations through multiple flows. 
Typical examples include constraint violations of application programming interfaces (APIs), active session attacks, web path traversal, and data harvesting.
Behavior-level attacks are generally detected through log data analysis, but this approach inherently suffers from response latency issues~\cite{luo2019botgraph,  prinakaa2024real}.
Because behavior-level attacks are increasing in prevalence and coexist with flow-level attacks, a traffic anomaly detection system must recognize not only flow-level but also behavior-level attacks.

To fill this gap, we propose BLADE, a \underline{B}ehavior-\underline{L}evel \underline{A}nomaly \underline{DE}tection system that relies on multi-flow traffic patterns in web services.
Our key observation is that web services typically expose a finite set of operations through their application-layer interfaces.
These operations exhibit distinctive communication patterns at the network layer, and these patterns can be captured from a multi-flow perspective, enabling differentiation of network flows generated by different web operations.

BLADE is a fully unsupervised and adaptive system.
It requires only unlabeled benign traffic data for an unsupervised training process and can quickly adapt to various web systems. 
BLADE consists of four core components: 
1) A flow autoencoder maps individual network flows into latent feature representations and generates per-flow reconstruction losses;
2) A pseudo operation label assignment component applies unsupervised clustering to discover web operational patterns and assign a pseudo operation label to each flow;
3) An anomaly score estimation component evaluates the anomaly degree of each flow using empirical cumulative distribution functions (ECDFs) on reconstruction losses; 
4) A multi-flow anomaly detection component aggregates the triplets of timestamps, pseudo labels, and anomaly scores from consecutive flows, then employs a feature extractor as well as a one-class support vector machine (OCSVM) for traffic anomaly detection.

The contributions of this work are summarized as follows.
\begin{itemize}
\item This paper is among the first to allow both flow-level and behavior-level attack detection in web services using network traffic.
The enabling observation is that application-layer operations of web services exhibit distinctive communication patterns at the network layer from a multi-flow perspective.

\item This paper proposes a novel traffic anomaly detection system, BLADE, that is fully unsupervised and adaptive.
The BLADE system constitutes a set of novel schemes, including pseudo operation label assignment, anomaly score estimation, and multi-flow anomaly detection. 

\item BLADE is evaluated on both custom and public datasets, demonstrating BLADE's superior performance with an F1 score of more than 0.97, outperforming state-of-the-art baselines and showing its effectiveness in detecting flow-level and behavior-level attacks in web services.

\end{itemize}

\section{Threat model}

\begin{figure}[t]
\centering
\includegraphics[width=\columnwidth]{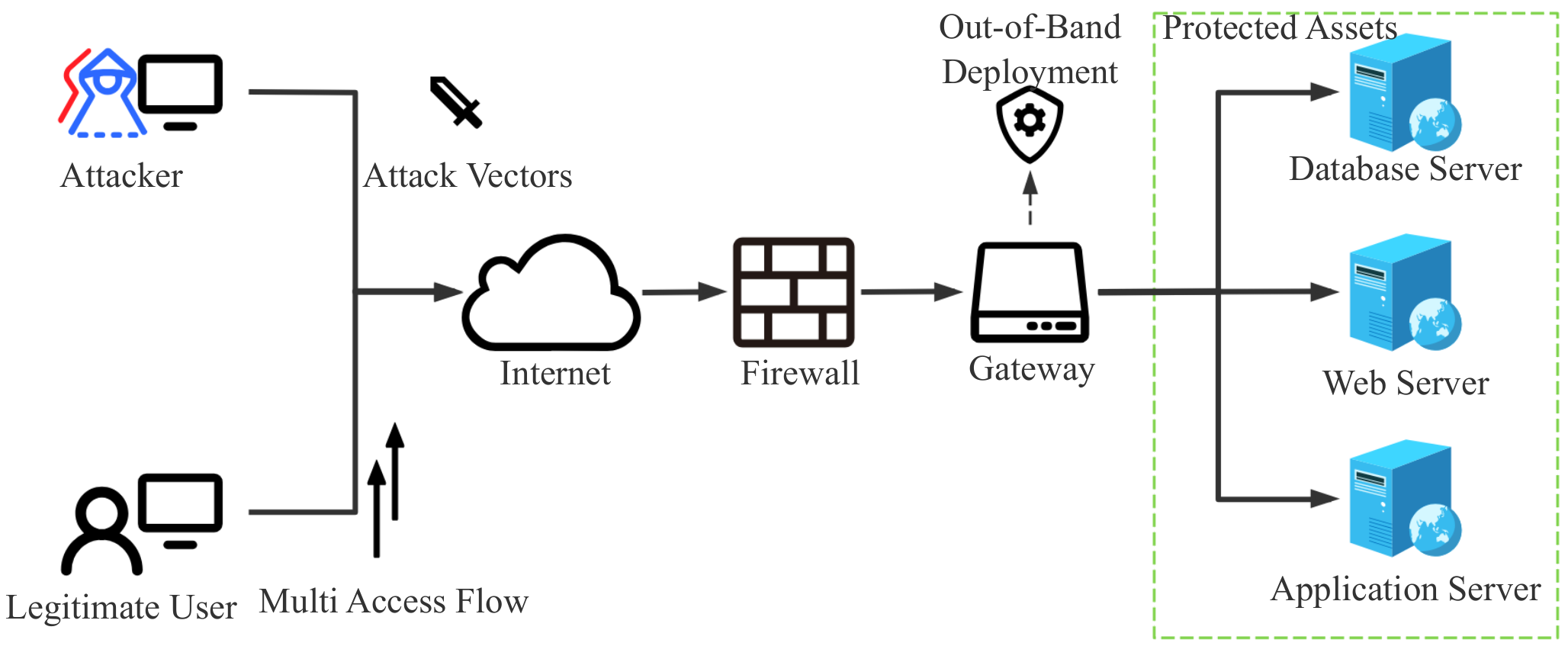}
\caption{Threat model.}
\label{ThreatModel}
\end{figure}

The threat model considered in this paper is shown in Fig.\ref{ThreatModel}.

\textbf{Protected Assets.} 
In this paper, protected assets are the computer systems that provide web services, facilitating communication between user clients and servers. 
Typically, these assets consist of web servers, application servers, database servers, and other supporting network infrastructures, which collaboratively ensure the seamless delivery of web services.
The protected assets can be viewed as one or a set of IP addresses and ports in cyberspace.

\textbf{Legitimate Users.}
A legitimate user can be represented as an IP address that accesses the protected assets through the Internet.
A single transmission control protocol (TCP) connection initiated by a user client to the protected assets is considered one access action, and the traffic generated by this TCP connection is referred to as a network flow.
An access action can be viewed as an operation on the protected assets.
When interacting with a web system, a user client typically performs continuous access actions. 
We refer to the sequence of actions within a time window as an access behavior, and these actions often exhibit contextual dependencies.
For instance, a user might first perform authentication through a login page, then navigate to a dashboard, subsequently access specific resources, and finally perform logout operations.
These sequential actions form coherent behavioral patterns that reflect legitimate usage workflows.

\textbf{Attackers and Attack Vectors.} 
The attackers are external individuals or groups that come from outside the protected assets and aim to compromise the security of the web system. 
Their objectives typically include disrupting web services, stealing sensitive information, or gaining unauthorized access.
These attackers can trigger both flow-level and behavior-level attacks.
In flow-level attacks, an attack vector is contained in a single flow.
Typical flow-level attacks comprise injection vectors exploiting application logic to compromise databases, denial-of-service (DoS) attacks disrupting service availability, botnet-driven attack automation, and reconnaissance scanning targeting vulnerability enumeration.
However, behavior-level attack vectors orchestrate multiple benign-appearing flows that collectively compromise system security.
This category encompasses adversarial scraping operations, including data harvesting, fraudulent bulk transactions, and resource hijacking, resulting in systemic service degradation and illegitimate infrastructure access.


\textbf{Defense Objectives.}
To defend against flow-level and behavior-level attacks, we propose a novel traffic anomaly detection technology called BLADE.
BLADE is deployed out-of-band at the web service provider's network gateway. 
It monitors network traffic via port mirroring, leaving the benign data flow intact.
Taking an access behavior as a basic detection unit, BLADE expands the capabilities of existing malicious traffic detection methods by simultaneously detecting both flow-level and behavior-level attacks in web services. 

\begin{figure*}[t]
\centerline{\includegraphics[width=.99\textwidth]{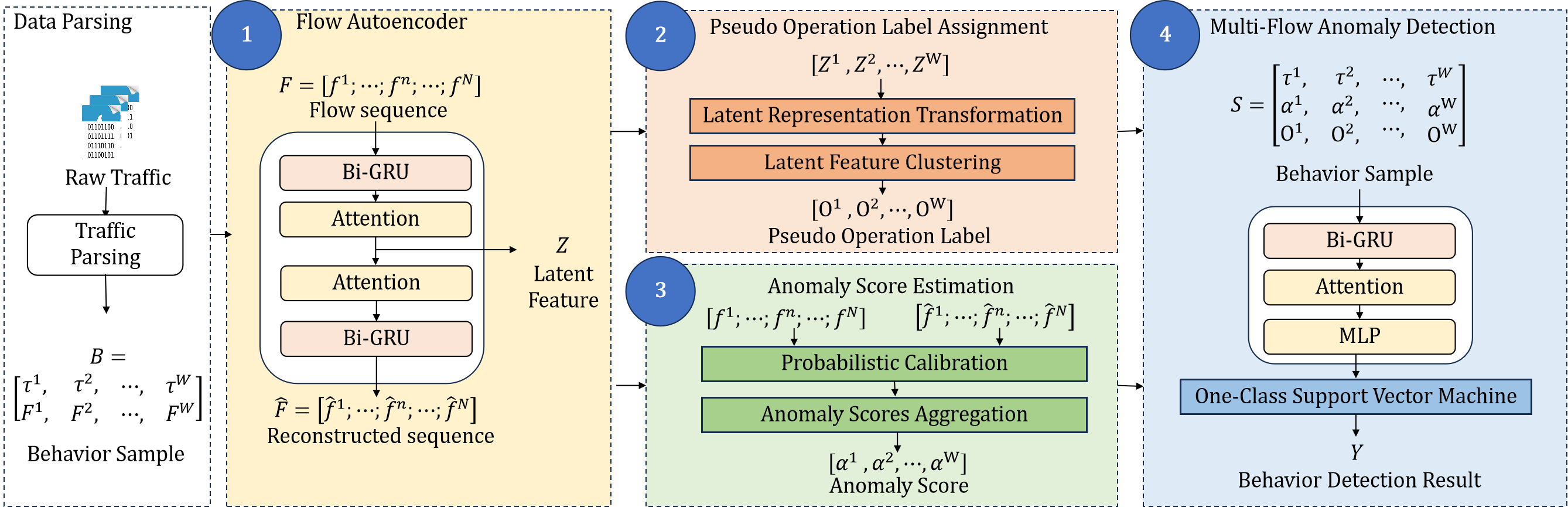}}
\caption{Framework of BLADE.}
\label{frammework}
\end{figure*}

\section{System Design of BLADE}
\subsection{System Overview}
This paper presents BLADE, a novel traffic anomaly detection framework that identifies not only flow-level but also behavior-level attacks in web services.
As shown in Fig.~\ref{frammework}, BLADE consists of four key components -- \textit{Flow Autoencoder}, \textit{Pseudo Operation Label Assignment}, \textit{Anomaly Score Estimation}, and \textit{Multi-Flow Anomaly Detection}.
Initially, the flow autoencoder maps the feature sequence of each flow into a latent feature representation and generates a reconstructed feature sequence.
Next, each latent feature representation is assigned a pseudo operation label using an unsupervised clustering algorithm.
The anomaly score is then calculated using the difference between the original and reconstructed feature sequence per flow.
Finally, the triplets of timestamps, pseudo operation labels, and anomaly scores from consecutive flows are transformed by a feature extractor, followed by a one-class classifier for anomaly detection.
The main advantages of BLADE are that it only requires unlabeled benign traffic data to automatically learn legitimate traffic patterns in an unsupervised manner.
Thus, BLADE is fully unsupervised and adaptive, and can be easily deployed in various web systems. 

\subsection{Data Parsing}\label{AA}
The first step of BLADE is to extract multi-flow samples, i.e., the basic units for anomaly detection, to characterize traffic patterns when a legitimate user accesses the protected assets.

\textbf{Multi-Flow Sample Extraction.}
For each captured network flow, we first extract $ N $ feature vectors, each of which consists of one type of packet-level attributes extracted from this flow.
Then, each feature vector is resized to a fixed length of $L$ by reserving the first $L$ values when longer or padding it with zeros when shorter.
Therefore, the feature sequence $F \in \mathbb{R}^{N \times L}$ extracted from each flow can be represented as
\begin{equation}
\begin{aligned}
F = \begin{bmatrix}
f^1; \cdots; f^n; \cdots; f^N
\end{bmatrix}.
\end{aligned}
\end{equation}
Therein, $ f^n \in \mathbb{R}^{1 \times L} $ is a fixed-length feature vector that encodes the $ n $-th attributes in the $ L $ consecutive packets.  
Moreover, each flow is associated with a first-seen timestamp $\tau \in \mathbb{R}\ge 0 $.


Since an access behavior consists of multiple network flows, we aggregate $W$ consecutive flows of an individual user in timestamp order into a multi-flow sample $B$. 
Specifically, a non-overlapping window with a size of $W$ is applied on a series of feature sequences extracted from each user's traffic.
In this way, a multi-flow sample $ B $ can be represented as
\begin{equation}
\begin{aligned}
B = \begin{bmatrix}
{\tau}^1 ,  \cdots ,{\tau}^w ,  \cdots , {\tau}^W \\
F^1 ,  \cdots ,F^w ,  \cdots , F^W \\
\end{bmatrix}.  \quad \\
\end{aligned}
\label{raw_B}
\end{equation}
Here, ${\tau}^w$ represents the timestamp of the $w$-th flow and $F^w$ denotes its feature sequence.
As indicated in Eq.~\eqref{raw_B}, the multi-flow sample $B$ contains packet-level attributes of multiple flows, enabling BLADE to detect both flow-level and behavior-level attacks.

\textbf{Training Dataset Construction.}
In anomaly detection, only benign traffic is collected and used during the training phase. 
We define $ \mathcal{B}$ as our training dataset. 
Therefore, $ \mathcal{B}$ can be represented as
\begin{equation}
\begin{aligned}
\mathcal{B} = \left\{ B^1, \cdots, B^m, \cdots, B^M  \right\},
\end{aligned}
\end{equation}
where $ B^m $ represents the $ m $-th multi-flow sample and $M$ represents the total number of training samples.

\subsection{Flow Autoencoder}
After extracting a multi-flow sample, BLADE exploits a customized autoencoder to learn the latent feature representation and reconstruction losses of each network flow.
The flow autoencoder has an encoder and a decoder.

\textbf{Encoder Architecture.}
The encoder is designed to learn a latent feature representation of each feature sequence $F$, using a bidirectional gated recurrent unit (BiGRU) followed by a self-attention layer.
The BiGRU processes the feature sequence $F$ along the time step. 
Each feature vector $f^n$ corresponds to a hidden state $h^n$ in the BiGRU output. 
The BiGRU learns the temporal relationships within each feature vector by processing it in both forward and backward directions, capturing dependencies from both past and future context within each flow. 
These hidden states are then concatenated to form the BiGRU output $H$, which is expressed as $H = \text{BiGRU}(F).$

Following the BiGRU, a self-attention layer is performed on $H$.
The attention result is further added to $H$.
A residual connection is applied between the original BiGRU output $H$ and the $H$ processed by the attention mechanism.
Then, a multi-layer perceptron (MLP) is leveraged to produce the final latent feature representation $Z$ as $Z = \text{MLP}(H + \text{Attention}(H)).$

\textbf{Decoder Architecture.}
The decoder reconstructs the feature sequence $F$ from the latent feature representation $Z$ using time-aware upsampling. 
The decoder is designed as the reverse process of the encoder, ensuring that the spatial and temporal dimensions of the latent feature representation are effectively upsampled to their original form.
For simplicity, we denote $\text{Decoder}(\cdot)$ as the entire decoding process.
It generates a reconstructed version of $F$ as
\begin{equation}
\begin{aligned}
\hat{F} &= \text{Decoder}(Z) = 
\begin{bmatrix}
\hat{f}^1; \cdots ; \hat{f}^n; \cdots; \hat{f}^N
\end{bmatrix},
\end{aligned}
\end{equation}
where $\hat{F}\in \mathbb{R}^{N \times L}$ represents the reconstructed feature sequence and $\hat{f}^n \in \mathbb{R}^{1 \times L} $ is reconstructed version of $f^n $.

\textbf{Reconstruction Loss.}
The reconstruction loss is calculated for each feature vector in $F$.
Given the feature vector $f^n$ in the $n$-th channel of $F$, we compute its mean squared error (MSE) between the original and reconstructed vectors as
\begin{equation}
\begin{aligned}
L^n = \frac{1}{L} \| f^{n} - \hat{f}^{n} \|_2^2,
\end{aligned}
\end{equation}
where $\| \cdot \|_2$ denotes the L2 norm.
Note that, $L^n$ will be used as the input data for the subsequent anomaly score estimation.


\textbf{Autoencoder Training.}
During the training phase, the flow autoencoder utilizes all feature vectors in the training dataset $ \mathcal{B} $ to capture the latent feature representations of legitimate network flows. 
Thus, the training loss of the flow autoencoder can be represented as
\begin{equation}
\begin{aligned}
\mathcal{L}^{\text{auto}} = \frac{1}{M} \sum_{m=1}^{M} \frac{1}{N} \sum_{n=1}^{N} L^n_{m},
\end{aligned}
\end{equation}
where $L^n_{m}$ denotes the MSE of the $n$-th attribute in the $m$-th flow.
In the detection phase, given a new multi-flow sample $B$, the flow autoencoder processes all feature vectors within it, producing corresponding latent feature representations and MSE losses.

\subsection{Pseudo Operation Label Assignment}
This component takes a latent feature representation $Z$ as input and assigns a pseudo operation label to it.
The reason behind this is that application-layer data, i.e., payload, is typically encapsulated into a network-layer packet using certain encryption schemes; thus, it is difficult to determine an exact operation when a user accesses web services based on encrypted network traffic.
Despite this, different user operations exhibit unique communication patterns that correspond to distinguishable latent representations, which allows operational pattern discovery via an unsupervised clustering method.

\textbf{Latent Feature Transformation.} 
Before feature clustering, we apply a series of transformations on each latent feature representation to enhance clustering robustness.
First, we perform low-variance filtering on $Z$ to eliminate uninformative feature dimensions. For each dimension across all training samples, we compute its empirical variance and retain only those dimensions with variance above a predefined threshold $\theta$, effectively removing features that show little variation and contribute minimal discriminative information.
Next, to reduce feature dimensionality while decorrelating the remaining features, the principal components analysis (PCA) with the whitening operation is applied.
We retain the smallest components that collectively capture at least 95\% of the total variance.
These transformations can generate compact and uncorrelated latent feature representations, thus improving clustering performance.
We use $Z'$ to represent the filtered version of $Z$, and $\text{PCA}_{\text{white}}(\cdot)$ to denote the PCA whitening process. 
$Z_w$ represents the latent representation after the above transformations, which can be expressed as
\begin{equation}
\begin{aligned}
Z_w = \text{PCA}_{\text{white}}(Z').
\end{aligned}
\end{equation}

\textbf{Latent Feature Clustering.}
In practice, the types of legitimate user operations are hard to determine based on network traffic, and different web services have different operations.
Under these conditions, an unsupervised clustering method is needed.
To meet this need, the hierarchical density-based spatial clustering of applications with noise (HDBSCAN) is selected to perform clustering analysis on the latent feature representations. 
HDBSCAN is a density-based clustering algorithm that automatically identifies clusters of noisy input features without a predefined cluster number.
Specifically, HDBSACN is first performed on all transformed latent feature representations, and the centers and boundaries of all clusters are determined.
For each transformed latent feature representation $Z_w$, HDBSCAN assigns a pseudo operation label $O$ as
\begin{equation}
\begin{aligned}
O = \text{HDBSCAN}(Z_w).
\end{aligned}
\end{equation}
It is worth noting that the pseudo operation label $O$ is just a cluster ID and corresponds to a certain web operation in the application layer.

\subsection{Anomaly Score Estimation}
This component takes the reconstruction losses between $F$ and $\hat{F}$ as input and outputs an anomaly score that indicates the anomaly degree of each network flow.

In the feature sequence $F$, different feature vectors have different value ranges and units.
This fact could result in some feature vectors dominating the overall reconstruction loss of the entire flow.
To address this issue, BLADE evaluates the anomaly degree of a new flow by computing its upper-tail probability in the historical distribution of reconstruction losses of each type of feature vectors in the training dataset. 
This ensures that different anomaly scores are balanced and normalized to the same range. 
Finally, all anomaly scores are aggregated into an overall anomaly score for this flow.

\textbf{Probabilistic Calibration.} 
Let $\mathbf{L}^{n} $ denote the collection of reconstruction losses belonging to the $n$-th attribute in all training feature sequences. 
The calibration process transforms these MSE loss values into probabilistic anomaly scores.
To do this, the empirical cumulative distribution function for each channel is constructed as 
\begin{equation}
\begin{aligned}
\text{ECDF}^{n}(x) = \frac{1}{|\mathbf{L}^{n}|} \sum_{i=1}^{|\mathbf{L}^{n}|} \mathbb{I}[\log(L_i^{n} + \epsilon) \leq x],
\end{aligned}
\end{equation}
where $ |\mathbf{L}^{n}|$ denotes the total number of training losses, $x$ is the query point, i.e., a log-transformed loss value, $L_i^{n}$ represents the $i$-th reconstruction loss for the $n$-th attribute in the training dataset, $\epsilon$ is a small constant to avoid zeros, and $\mathbb{I}[\cdot]$ is the indicator function, which outputs the value 1 when the condition is satisfied, and 0 otherwise.
Moreover, the log-transformation is applied to smooth and standardize reconstructed losses.

During testing, given a new reconstruction loss $L_{\text{test}}^{n}$, the anomaly score is computed as the negative value of the log-transformation of its upper-tail probability
\begin{align}
p^{n} &= 1 - \text{ECDF}^{n}(\log(L_{\text{test}}^{n} + \epsilon)), \\
a^{n} &= -\log(p^{n} + \delta),
\end{align}
where $p^{n}$ represents the upper-tail probability for the $n$-th attribute, $a^{n}$ is the corresponding anomaly score, and $\delta$ is used to prevent zero probabilities. 
A higher anomaly score indicates a greater deviation from the benign traffic pattern.

\textbf{Anomaly Score Aggregation.} 
To derive a final anomaly score for each flow, we use the LogSumExp function to aggregate the anomaly scores computed across all attributes as
\begin{equation}
\begin{aligned}
\alpha = \log\left(\sum_{n=1}^{N} \exp(a^n)\right).
\end{aligned}
\end{equation}
This function is similar to the softmax function and places greater emphasis on larger anomaly scores while reducing the influence of smaller ones.
The result is a final anomaly score $\alpha$ that reflects the overall anomaly degree of this flow.
Given the reconstruction losses of all flows in the multi-flow sample $B$, this component calculates their anomaly scores, providing a measure of deviation from legitimate behavior patterns.

\subsection{Multi-Flow Anomaly Detection}
For each multi-flow sample $B$, this component first aggregates its pseudo operation labels and anomaly scores into a behavior sample $S$.
Then, a feature extractor encodes $S$ into a latent feature representation, and a one-class support vector machine is exploited to detect traffic anomalies.

First, a behavior sample $S$ can be expressed as
\begin{equation}
\begin{aligned}
S = \begin{bmatrix}
{\tau}^1,\cdots, {\tau}^w , \cdots, {\tau}^W \\
 \alpha^1 , \cdots , \alpha^w , \cdots , \alpha^W \\
 O^1 , \cdots, O^w , \cdots ,  O^W
\end{bmatrix},
\end{aligned}
\end{equation}
where \( \tau^w \), \( \alpha^w \), and \( O^w \) represent the timestamp, anomaly score, and pseudo operation label of the \( w \)-th flow in the multi-flow sample $B$, respectively.
Then, we use the behavior samples extracted from the training dataset $ \mathcal{B} $ to train an autoencoder, which has the same structure as the flow autoencoder.
Next, we take its encoder part as a feature extractor.
This design aims to capture the contextual dependencies within $S$.
The output of the extractor $X$ is denoted as
\begin{equation}
\begin{aligned}
X =\text{Extractor}(S).
\end{aligned}
\end{equation}
The behavior representation $X$ is then passed to a OCSVM. 

The OCSVM is used to distinguish between legitimate and malicious traffic by building the decision boundary of representations of benign data. 
Since this method works in an unsupervised manner, it does not require labeled anomalous data in the training phase.
During the detection phase, the learned OCSVM processes the behavior representation and outputs a detection result $Y$ as
\begin{equation}
\begin{aligned}
Y=\text{OCSVM}(X).
\end{aligned}
\end{equation}

In summary, BLADE is an unsupervised and adaptive traffic anomaly detection system that effectively detects both flow-level and behavior-level attacks in diverse web systems.
This approach enables efficient and robust anomaly detection in environments where labeled data is scarce or unavailable.

\section{Experimental Evaluation}
\subsection{Evaluation Methodology}
We evaluate BLADE on both our custom dataset and the publicly available CIC-IDS2017 dataset~\cite{sharafaldin2018toward}.

\textbf{Custom Dataset.}
To the best of our knowledge, no public dataset provides labeled traffic under flow-level and behavior-level attacks.
Hence, we build a reproducible blog-style web testbed with a fixed number of web operations.
DPDK\cite{dpdk} is used to capture network traffic via 10G optical modules for data acquisition.
To collect benign traffic, ten volunteers are recruited to normally interact with the web testbed using different clients.
In this condition, the traffic of each user’s session is captured.
To collect anomalous traffic, we launch both the flow-level and behavior-level attacks on the web testbed.
The flow-level attacks include DoS attacks, injection attacks, brute-force attacks, and scanning attacks.
As for behavior-level attacks, active session attacks, web bot bulk operations, and malicious data harvesting attacks are implemented. 
In this way, a total of 597296 benign traffic flows and 123474 malicious flows are collected.
For each flow, packet-level attributes, including packet size, inter-arrival time, and TCP control flags, are extracted.
In this way, this dataset contains 11965 benign multi-flow samples and 2480 malicious ones.

\textbf{CIC-IDS2017 Dataset.}
The CIC-IDS2017 \cite{sharafaldin2018toward} dataset is a well-known benchmark dataset for intrusion detection and provides labeled network traffic under benign conditions and multiple types of real-world attacks, such as DoS, port scans, brute force, botnet, and web attacks.
Several studies have reported some issues in the CIC-IDS2017 dataset, such as packet misordering, packet duplication, and mislabeling.
To deal with these issues, we adopt the correction methods suggested in the literature~\cite{cicidsfix, liu2022error}.
To generate multi-flow samples, the flows that have the same IP address and port number are considered to be from the same user and associated in chronological order. 
In this way, we yield 22753 benign multi-flow samples and 2642 malicious ones.
Because the CIC-IDS2017 dataset does not involve behavior-level attacks, this dataset is only used to evaluate BLADE’s detection performance on flow-level attacks. 

\textbf{Training and Testing.} 
We implement BLADE on a workstation equipped with an Intel Core i7-13700K CPU, NVIDIA GeForce RTX 4090 GPU, and 256 GB RAM, running Ubuntu 20.04.6. 
All components are developed using Python 3.12 and PyTorch 2.5.0. 
Table~\ref{tab:hyperparameters} lists the specifications of BLADE's hyperparameters in our experiment.
For model training and testing, the benign samples are partitioned into training and testing samples with a ratio of 7:3 in each of the two datasets.
Moreover, all malicious samples are used for testing.

\begin{table}[t]
\centering
\caption{Settings of BLADE's Hyperparameters in Our Experiment}
\label{tab:hyperparameters}
\resizebox{\columnwidth}{!}{
\begin{tabular}{llc}
\toprule
\textbf{Component} & \textbf{Hyperparameter} & \textbf{Value} \\
\midrule
\multirow{3}{*}{\textbf{Data Parsing}} 
& Length of Feature Vector ($L$) & 50 \\
& Size of Behavior Window ($W$) & 50 \\
& Number of Packet-level Attributes ($N$) & 3 \\
\midrule
\multirow{3}{*}{\textbf{Flow Autoencoder}} 
& Hidden State Size & 128 \\
& Latent Representation Dimension & 64 \\
& Number of BiGRU Layers & 2 \\
\midrule
\multirow{1}{*}{\textbf{Pseudo Operation Label Assignment}}
& Low-Variance Threshold ($\theta$) & 0.01 \\
\midrule
\multirow{2}{*}{\textbf{Anomaly Score Estimation}}
& Log-Transformation Parameter ($\epsilon$) & $10^{-8}$ \\
& Probability Parameter ($\delta$) & $10^{-8}$ \\
\bottomrule
\end{tabular}
}
\end{table}

\textbf{Evaluation Metrics.} 
We use the following metrics to evaluate BLADE's performance.
\begin{itemize}
\item \textbf{Precision.} It is the ratio of correctly identified malicious samples to all samples predicted as malicious.
\item \textbf{Recall.} It is the ratio of correctly identified malicious samples to all actual malicious samples.
\item \textbf{F1 Score.} It is the harmonic mean of precision and recall, providing a balanced measure of system performance.
\item \textbf{Silhouette Coefficient.} It is the similarity of an object to its own cluster compared to other clusters.
\item \textbf{Calinski-Harabasz Index.} It is a metric that evaluates the ratio of the sum of inter-cluster dispersion to that of intra-cluster dispersion. 
\item \textbf{Davies-Bouldin Index.} It is based on the average similarity ratio of each cluster to its most-similar cluster. 
\end{itemize}
The first three metrics measure the anomaly detection performance of BLADE, while the latter three are used to evaluate the performance of unsupervised clustering involved in BLADE.

\subsection{Experimental Results}

\begin{table}[t]
\centering
\caption{Performance of BLADE on Custom Dataset}
\label{tab:workflow_dataset_results}
\resizebox{\columnwidth}{!}{
\begin{tabular}{l l c c c}
\Xhline{1.5pt} 
Attack Type & Attack Vector  & Precision  & Recall & F1 score  \\
\Xhline{1.5pt}
\multirow{4}{*}{Flow-Level} & DoS & 0.9661 & 0.9975 & 0.9816 \\
                                & Scan & 0.9296 & 0.9965 & 0.9619  \\
                                & Injection & 0.9708 & 0.9677 & 0.9693 \\
                                & Brute Force & 0.9759 & 0.9179 & 0.9460 \\
\hline
\multirow{3}{*}{Behavior-Level} & Data Harvesting & 0.9661 & 0.9907 & 0.9782 \\
                                & Active Session & 0.9894 & 0.9993 & 0.9943  \\
                                & Web Botnet Operation & 0.9635 & 1.0000 & 0.9814 \\
\Xhline{1.5pt} 
\multirow{1}{*}{Average}
& - & 0.9659 & 0.9814 & 0.9732 \\
\Xhline{1.5pt} 
\end{tabular}
}
\end{table}

\textbf{Anomaly Detection on Custom Dataset.}
First, we present BLADE's anomaly detection performance on the custom dataset.
As shown in Table~\ref{tab:workflow_dataset_results}, BLADE achieves consistently high performance against the four flow-level attacks.
Especially, it obtains a precision of 0.9661, an exceptional recall of 0.9975, and a high F1 score of 0.9816 for DoS attacks.
As for behavior-level attacks, BLADE has the best performance when detecting active session attacks, with a precision of 0.9894, a high recall of 0.9993, and a near-perfect F1 score of 0.9943. 
On average, BLADE achieves a precision of 0.9659, a recall of 0.9814, and an F1 score of 0.9732, indicating its high effectiveness in recognizing both flow-level and behavior-level web service attacks.

\begin{table}[t]
\centering
\caption{Performance of BLADE on CIC-IDS2017 Dataset}
\label{tab:table_cicids2017_results}
\begin{tabular}{c c c c}
\Xhline{1.5pt} 
Attack Vector  & Precision  & Recall & F1 score  \\
\Xhline{1.5pt} 
DoS & 0.9661 & 0.9907 & 0.9782 \\
Botnet & 0.9894 & 0.9993 & 0.9943  \\
Port Scan & 0.9991 & 1.0000 & 0.9995 \\
Web Attack & 0.9214 & 0.9744 & 0.9472  \\
Brute Force & 0.9635 & 1.0000 & 0.9814 \\
\Xhline{1.5pt} 
\multirow{1}{*}{Average}
& 0.9679 & 0.9929 & 0.9801 \\
\Xhline{1.5pt} 
\end{tabular}
\end{table}

\textbf{Anomaly Detection on CIC-IDS2017 Dataset.}
Then, we report BLADE's performance on the CIC-IDS2017 dataset for detecting flow-level attacks in Table~\ref{tab:table_cicids2017_results}.
BLADE achieves the highest F1 score of 0.9995 with a perfect recall for port scan detection.
As for Botnet attacks, our system presents an F1 score of 0.9943 and a high recall of 0.9993.
The system obtains the lowest performance against web attacks with a precision of 0.9214. 
Despite that, BLADE achieves a precision of 0.9679, a recall of 0.9929, and an F1 score of 0.9801 on average, validating its robustness across various flow-level attacks.

\begin{figure}[t]
\centering
\resizebox{\columnwidth}{!}{\includegraphics{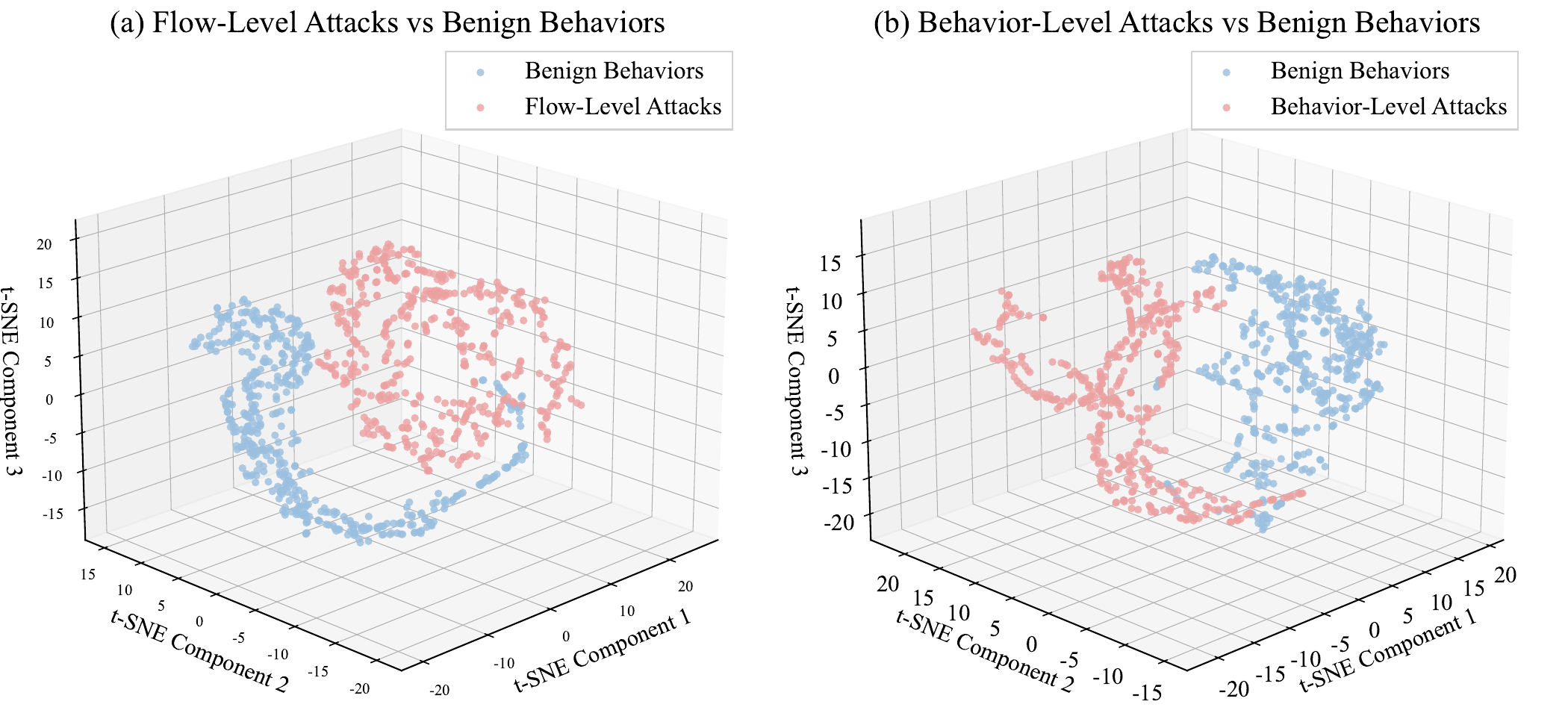}}
\caption{t-SNE visualization of malicious and benign traffic.}
\label{fig:behavior_embedding_visualization}
\end{figure}

\textbf{Behavioral Embedding Visualization.}
To highlight the differences between malicious and benign traffic, we apply t-distributed stochastic neighbor embedding (t-SNE) to the samples from our custom dataset.
Fig.~\ref{fig:behavior_embedding_visualization} visualizes high-dimensional embeddings from the multi-flow anomaly detection module.
Overall, the projection results show high distinguishability between benign and malicious traffic in the latent space.
In Fig.~\ref{fig:behavior_embedding_visualization}~(a), flow-level attacks are clearly separated from benign ones, indicating that triplet-sequence modeling is effective in capturing temporal and contextual cues.
In Fig.~\ref{fig:behavior_embedding_visualization}~(b), the latent embeddings of behavior-level attacks are distant from those of benign traffic.
The strong separability stems from distinctive multi-flow operation sequences (e.g., bulk operations, data harvesting) that are invisible per flow but revealed by our pseudo-operation labels and anomaly-score integration.
These clear discriminating boundaries confirm that BLADE can distinguish both flow-level and behavior-level anomalies within a unified latent space of multi-flow features.

\begin{figure}[t]
\centering
\resizebox{\columnwidth}{!}{\includegraphics{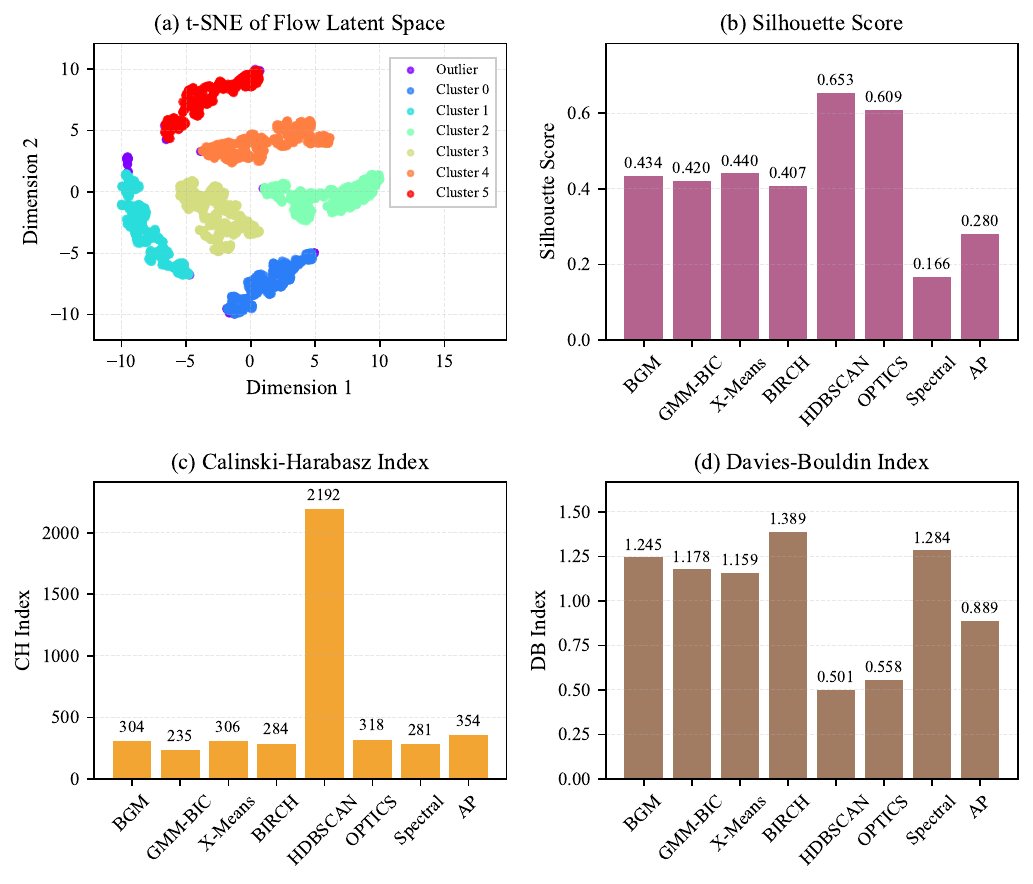}}
\caption{Performance of different clustering algorithms.}
\label{fig:clustering_performance_comparison}
\end{figure}

\textbf{Impact of Cluster Algorithms.}
We evaluate eight representative clustering algorithms covering four categories: 1) probabilistic model-based methods, including Bayesian Gaussian mixture (BGM) and Gaussian mixture model with Bayesian information criterion (GMM-BIC); 2) prototype-based methods, including X-means and balanced iterative reducing and clustering using hierarchies (BIRCH); 3) density-based methods, consisting of HDBSCAN and ordering points to identify the clustering structure (OPTICS); 4) graph-based approaches, including spectral clustering with eigenvector (SCE) and affinity propagation (AP).
Fig.~\ref{fig:clustering_performance_comparison}~(a) illustrates the latent embeddings derived from benign traffic in our custom dataset.
Six non-overlapping clusters can be clearly observed.
These clusters correspond to different application-layer operations, which lay the foundation for applying clustering algorithms to distinguish differences in operations.
According to Fig.~\ref{fig:clustering_performance_comparison}~(b)-(d), HDBSCAN achieves the best clustering performance, with the highest Silhouette score of 0.653, the lowest Davies-Bouldin index of 0.501, and the highest Calinski-Harabasz index of 2192. 
HDBSCAN's superior performance stems from its density-based approach that effectively discovers clusters with different point densities, making it well-suited for grouping feature points of benign traffic, as shown in Fig.~\ref{fig:clustering_performance_comparison}~(a).
These results justify the adoption of HDBSCAN in our system for latent feature clustering.

\begin{figure}[t]
\centering
\includegraphics[width=\columnwidth]{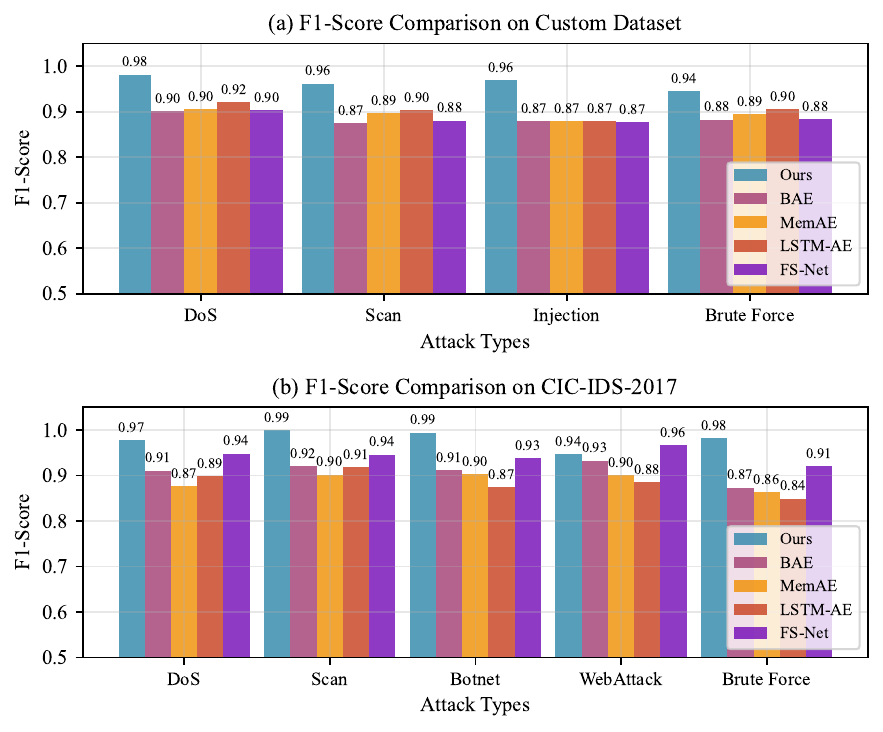}
\caption{Performance comparison between BLADE and flow-level baselines on the custom and CIC-IDS2017 datasets.}
\label{fig:f1_score_comparison}
\end{figure}

\textbf{Comparison with Single-Flow Baselines.}
Because no existing approach exploits network traffic for detecting behavior-level attacks in web services, we conduct a comparative study against four representative single-level baselines for flow-level attack detection.
The selected baselines include BAE~\cite{wang2023bae}, MemAE~\cite{min2021memae}, LSTM-AE~\cite{lstmae}, and FS-Net~\cite{fsnet}.
We evaluate their anomaly detection performance on both the custom and CIC-IDS2017 datasets.
As presented in Fig.~\ref{fig:f1_score_comparison}, BLADE shows superior performance on most of the attack vectors in the two datasets.
As depicted in Fig.~\ref{fig:f1_score_comparison}~(a), the F1 score of BLADE is at least 0.09 higher than that of the baselines in injection attacks and 0.04 higher than that in brute force attacks on the custom dataset.   
When it comes to the CIC-IDS-2017 dataset, FS-Net achieves the best performance among the four baselines, and even has a higher F1 score in web attacks than BLADE.
However, BLADE outperforms FS-Net in the other four types of attacks.
These results validate the superiority of multi-flow anomaly detection over traditional single-flow approaches. 


\textbf{Ablation Study.}
Finally, we conduct an ablation study to validate the effectiveness of each component in BLADE.
For this purpose, we build three variants by systematically removing one component from BLADE at a time: 
1) The first variant removes the anomaly score estimation process and directly feeds raw reconstruction losses into multi-flow anomaly detection with pseudo operation labels and timestamps; 
2) The second variant excludes pseudo operation labels, utilizing only anomaly scores and timestamps for anomaly detection; 
3) The third variant removes anomaly scores, employing only pseudo operation labels and timestamps in multi-flow anomaly detection.
We evaluate the three variants on our custom dataset and report their anomaly detection results in Table~\ref{tab:ablation_results}.
As the table shows, the precisions, recalls, and F1 scores of all variants are at least 0.32, 0.36, and 0.34 lower than those of BLADE, respectively.
Specifically, the first variant has the worst performance, with a low precision of 0.3414.
This indicates that the raw reconstruction losses are of little help in anomaly detection.
The second variant suffers a precision decrease of more than 0.5, suggesting that pseudo operation labels are helpful in characterizing meaningful patterns of legitimate web service traffic.
The third variant obtains the highest performance among all variants, but still presents a significant performance degradation.
The above results show the high effectiveness of each component proposed in BLADE.

\begin{table}[t] 
\centering
\caption{Ablation Study of BLADE}
\label{tab:ablation_results}
\resizebox{\columnwidth}{!}{
\begin{tabular}{llcccc}
\toprule
Variants & Precision  & Recall & F1 Score \\
\midrule
\#1: w/o Anomaly Score Estimation & 0.3414 & 0.5432 & 0.4193 \\
\#2: w/o Pseudo Labels& 0.4574 & 0.6428 & 0.5345 \\
\#3: w/o Anomaly Scores& 0.6386 & 0.6157 & 0.6269 \\
\hline
BLADE (Complete Model)                        & 0.9659 & 0.9814 & 0.9732 \\
\bottomrule
\end{tabular}}
\end{table}

\section{Related work}
\textbf{Traffic Anomaly Detection.}
Traffic anomaly detection is widely adopted to protect the security of web services.
Liu et al.~\cite{fsnet} proposed the flow sequence network (FS-Net), which extracts representative features from raw network flows and classifies them from a single-flow perspective.
Qing et al.~\cite{qing2023low} proposed RAPIER that leverages the distinct distributions of legitimate and malicious traffic flows in the feature space to augment training data.
Additionally, there are some methods based on mixed traffic, where a burst refers to a significant and sudden increase in the flow size within a given time window, without the need to classify individual flows.  
Cheng et al.~\cite{chengburst} proposed BurstDetector, incorporating the definition of across-period bursts and employing a two-stage detection process.
However, these approaches mainly focus on single-flow characteristics and fail to capture behavioral-level attacks that span multiple flows.
In contrast, BLADE is the first traffic anomaly detection system that exploits multi-flow features and is capable of detecting not only flow-level but also behavior-level attacks.

\textbf{User Behavior Analysis.}
User behavior analysis is also used to secure web services.
Luo et al.~\cite{luo2019botgraph} proposed BotGraph, using a pre-obtained sitemap to convert user behaviors from log data into subgraphs for behavior classification.
Prinakaa et al.~\cite{prinakaa2024real} proposed a real-time API abuse detection system that utilizes behavioral analysis of API logs to identify and mitigate security threats.
Because log data is hard to access in many cases, researchers are beginning to investigate the feasibility of user behavior analysis using network traffic. 
Mengmeng et al.~\cite{mengmeng2025enmob} proposed Enmob, a multi-flow-based behavioral traffic classification method, which is designed to uncover application behaviors using encrypted application traffic.
However, this method requires predefined behavior patterns and labeled traffic. 
Differently, BLADE focuses on anomaly detection and can automatically generate pseudo operation labels in an unsupervised manner.

\section{Conclusion}

This paper presents BLADE, a novel traffic anomaly detection system that can detect both the flow-level and behavior-level attacks in web services.
We observe that application-layer operations of web services exhibit distinctive communication patterns at the network layer from a multi-flow perspective.
BLADE generates a pseudo operation label and an anomaly score for each flow and learns behavior patterns of legitimate web users using features within multiple flows, thus facilitating the detection of single-flow and multi-flow anomalies.
We implement BLADE and evaluate it on our custom and public CIC-IDS2017 datasets.
The evaluation results demonstrate that BLADE achieves an average F1 score of 0.9732 against both the flow-level and behavior-level attacks.
In addition, BLADE outperforms four flow-level baselines, showing the superiority of multi-flow anomaly detection over traditional single-flow approaches.

\section*{Acknowledgement}
This work was supported in part by the National Natural Science Foundation of China with Grant 62301499 and the Henan Association for Science and Technology with Grant 2025HYTP037.

\bibliography{ref}  

@article{qing2023low,
  title={Low-quality training data only? A robust framework for detecting encrypted malicious network traffic},
  author={Qing, Yuqi and Yin, Qilei and Deng, Xinhao and Chen, Yihao and Liu, Zhuotao and Sun, Kun and Xu, Ke and Zhang, Jia and Li, Qi},
  journal={arXiv preprint arXiv:2309.04798},
  year={2023}
}

@article{sharafaldin2018toward,
  title={Toward generating a new intrusion detection dataset and intrusion traffic characterization.},
  author={Sharafaldin, Iman and Lashkari, Arash Habibi and Ghorbani, Ali A and others},
  journal={ICISSp},
  volume={1},
  number={2018},
  pages={108--116},
  year={2018}
}

@misc{dpdk,
   title={Data Plane Development Kit ({DPDK})},
   author={Linux Foundation},
   year={2015},
   url = {http://www.dpdk.org}
}

@article{min2021memae,
  title={Network anomaly detection using memory-augmented deep autoencoder},
  author={Min, Byeongjun and Yoo, Jihoon and Kim, Sangsoo and Shin, Dongil and Shin, Dongkyoo},
  journal={IEEE Access},
  volume={9},
  pages={104695--104706},
  year={2021},
  publisher={IEEE}
}

@article{wang2023bae,
  title={Bae: Anomaly detection algorithm based on clustering and autoencoder},
  author={Wang, Dongqi and Nie, Mingshuo and Chen, Dongming},
  journal={Mathematics},
  volume={11},
  number={15},
  pages={3398},
  year={2023},
  publisher={MDPI}
}

@INPROCEEDINGS{fsnet,
  author={Liu, Chang and He, Longtao and Xiong, Gang and Cao, Zigang and Li, Zhen},
  booktitle={IEEE INFOCOM 2019 - IEEE Conference on Computer Communications}, 
  title={FS-Net: A Flow Sequence Network For Encrypted Traffic Classification}, 
  year={2019},
  volume={},
  number={},
  pages={1171-1179},
  doi={10.1109/INFOCOM.2019.8737507}}

@misc{akamaistate2025,
  title        = {State of Apps and API Security 2025: How AI is Shifting the Digital Terrain},
  author       = {Akamai Technologies},
  year         = {2025},
  howpublished = {\url{https://mysecuritymarketplace.com/reports/state-of-apps-and-api-security-2025-how-ai-is-shifting-the-digital-terrain/}},
  note         = {Accessed: 2025-09-14}
}

@inproceedings{lstmae,
  title={Network anomaly detection using LSTM based autoencoder},
  author={Said Elsayed, Mahmoud and Le-Khac, Nhien-An and Dev, Soumyabrata and Jurcut, Anca Delia},
  booktitle={Proceedings of the 16th ACM symposium on QoS and security for wireless and mobile networks},
  pages={37--45},
  year={2020}
}

@InProceedings{cicidsfix,
author="Lanvin, Maxime
and Gimenez, Pierre-Fran{\c{c}}ois
and Han, Yufei
and Majorczyk, Fr{\'e}d{\'e}ric
and M{\'e}, Ludovic
and Totel, {\'E}ric",
editor="Kallel, Slim
and Jmaiel, Mohamed
and Zulkernine, Mohammad
and Hadj Kacem, Ahmed
and Cuppens, Fr{\'e}d{\'e}ric
and Cuppens, Nora",
title="Errors in the CICIDS2017 Dataset and the Significant Differences in Detection Performances It Makes",
booktitle="Risks and Security of Internet and Systems",
year="2023",
publisher="Springer Nature Switzerland",
address="Cham",
pages="18--33",
isbn="978-3-031-31108-6"
}

@inproceedings{liu2022error,
  title={Error prevalence in nids datasets: A case study on cic-ids-2017 and cse-cic-ids-2018},
  author={Liu, Lisa and Engelen, Gints and Lynar, Timothy and Essam, Daryl and Joosen, Wouter},
  booktitle={2022 IEEE Conference on Communications and Network Security (CNS)},
  pages={254--262},
  year={2022},
  organization={IEEE}
}

@article{nascita2024survey,
  title={A survey on explainable artificial intelligence for internet traffic classification and prediction, and intrusion detection},
  author={Nascita, Alfredo and Aceto, Giuseppe and Ciuonzo, Domenico and Montieri, Antonio and Persico, Valerio and Pescap{\'e}, Antonio},
  journal={IEEE Communications Surveys \& Tutorials},
  year={2024},
  publisher={IEEE}
}

@article{dong2025deep,
  title={Deep learning and pre-training technology for encrypted traffic classification: A comprehensive review},
  author={Dong, Wenqi and Yu, Jing and Lin, Xinjie and Gou, Gaopeng and Xiong, Gang},
  journal={Neurocomputing},
  volume={617},
  pages={128444},
  year={2025},
  publisher={Elsevier}
}

@article{liu2025decentralized,
  title={Decentralized traffic detection utilizing blockchain-federated learning with quality-driven aggregation},
  author={Liu, Wei and Cui, Wentao and Wang, Bin and Pan, Heng and She, Wei and Tian, Zhao},
  journal={Computer Networks},
  volume={262},
  pages={111179},
  year={2025},
  publisher={Elsevier}
}

@article{fu2024flow,
  title={Flow interaction graph analysis: Unknown encrypted malicious traffic detection},
  author={Fu, Chuanpu and Li, Qi and Xu, Ke},
  journal={IEEE/ACM Transactions on Networking},
  volume={32},
  number={4},
  pages={2972--2987},
  year={2024},
  publisher={IEEE}
}

@inproceedings{Mateenad,
author = {Alotaibi, Fahad and Maffeis, Sergio},
title = {Mateen: Adaptive Ensemble Learning for Network Anomaly Detection},
year = {2024},
isbn = {9798400709593},
publisher = {Association for Computing Machinery},
address = {New York, NY, USA},
url = {https://doi.org/10.1145/3678890.3678901},
doi = {10.1145/3678890.3678901},
booktitle = {Proceedings of the 27th International Symposium on Research in Attacks, Intrusions and Defenses},
pages = {215–234},
numpages = {20},
location = {Padua, Italy},
series = {RAID '24}
}

@inproceedings{han2023anomaly,
  title={Anomaly Detection in the Open World: Normality Shift Detection, Explanation, and Adaptation.},
  author={Han, Dongqi and Wang, Zhiliang and Chen, Wenqi and Wang, Kai and Yu, Rui and Wang, Su and Zhang, Han and Wang, Zhihua and Jin, Minghui and Yang, Jiahai and others},
  booktitle={NDSS},
  year={2023}
}

@inproceedings{prinakaa2024real,
  title={A Real-Time Approach to Detecting API Abuses Based on Behavioral Patterns},
  author={Prinakaa, Sameeraa and Bavanika, V and Sanjana, S and Srinivasan, Sneha and Sarasvathi, V},
  booktitle={2024 8th International Conference on Cryptography, Security and Privacy (CSP)},
  pages={24--28},
  year={2024},
  organization={IEEE}
}

@INPROCEEDINGS{chengburst,
  author={Cheng, Zhongyi and Gao, Guoju and Huang, He and Sun, Yu-E and Du, Yang and Wang, Haibo},
  booktitle={IEEE INFOCOM 2024 - IEEE Conference on Computer Communications}, 
  title={BurstDetector: Real-Time and Accurate Across-Period Burst Detection in High-Speed Networks}, 
  year={2024},
  volume={},
  number={},
  pages={2338-2347},
  doi={10.1109/INFOCOM52122.2024.10621114}}

@article{luo2019botgraph,
  title={Botgraph: Web bot detection based on sitemap},
  author={Luo, Yang and She, Guozhen and Cheng, Peng and Xiong, Yongqiang},
  journal={arXiv preprint arXiv:1903.08074},
  year={2019}
}

@article{mengmeng2025enmob,
  title={Enmob: Unveil the Behavior with Multi-flow Analysis of Encrypted App Traffic},
  author={Mengmeng, Ge and Ruitao, Feng and Likun, Liu and Xiangzhan, Yu and Vinay, Sachidananda and Xiaofei, Xie and Yang, Liu},
  journal={Cybersecurity},
  volume={8},
  number={1},
  pages={26},
  year={2025},
  publisher={Springer}
}


\end{document}